\documentclass[12pt]{article}
\usepackage{epsfig}
\newcommand{\be}{\begin{equation}}
\newcommand{\ee}{\end{equation}}
\newcommand{\bq}{\begin{quote}}
\newcommand{\eq}{\end{quote}}
\begin{document}
\setlength{\baselineskip}{16.5pt}

\title{The Pondicherry interpretation of quantum mechanics: An overview}
\author{U. Mohrhoff\\
Sri Aurobindo International Centre of Education\\
Pondicherry 605002 India\\
\normalsize\tt ujm@auromail.net}
\date{}
\maketitle
\begin{abstract}
\noindent An overview of the Pondicherry interpretation of quantum mechanics is presented. This 
interpretation proceeds from the recognition that the fundamental theoretical framework of physics is 
a probability algorithm, which serves to describe an objective fuzziness (the literal meaning of 
Heisenberg's term ``Unsch\"arfe,'' usually mistranslated as ``uncertainty'') by assigning objective 
probabilities to the possible outcomes of unperformed measurements. Although it rejects attempts to 
construe quantum states as evolving ontological states, it arrives at an objective description of the 
quantum world that owes nothing to observers or the goings-on in physics laboratories. In fact, 
unless such attempts are rejected, quantum theory's true ontological implications cannot be seen. 
Among these are the radically relational nature of space, the numerical identity of the corresponding 
relata, the incomplete spatiotemporal differentiation of the physical world, and the consequent 
top-down structure of reality, which defies attempts to model it from the bottom up, whether on the 
basis of an intrinsically differentiated spacetime manifold or out of a multitude of individual building 
blocks.

\vspace{6pt}\noindent {\it Keywords\/}: interpretation of quantum mechanics; Pondicherry 
interpretation;

\noindent objective probabilities; fuzziness, space; time.

\vspace{6pt}\noindent {\it PACS numbers\/}: 03.65.Ta; 03.65.-w; 01.70.+w
\setlength{\baselineskip}{14pt}
\end{abstract}
\pagebreak

\section{\large Introduction}
Over the past five years, a new interpretation of quantum mechanics has emerged. After a series of 
publications that have focused on various aspects of the evolving ``Pondicherry interpretation''
[1--11], it is time for an overview.

Does quantum theory need an interpretation? In a wide-circulation opinion piece, Fuchs and 
Peres~\cite{FuPer} have claimed that it does not. Their actual claims, however, are that ``quantum 
theory does {\it not\/} describe physical reality'' (original emphasis), and that ``[t]he compendium 
of probabilities represented by the `quantum state' $\rho$ captures everything that can meaningfully 
be said about a physical system.'' I am in full agreement with these authors when they insist that a 
quantum state (including the ``two-state'' introduced by Aharonov and Vaidman~\cite{AV}) is 
nothing but a compendium of probabilities. The objective fuzziness of relative positions and momenta 
is essential for the stability of matter. An objective fuzziness requires objective probabilities for its 
description. And a quantum state is an algorithm for calculating objective probabilities. In fact, the 
mathematical structure of quantum theory follows directly from this characterization of quantum 
states~\cite{MohrhoffJustSo}.

However, I disagree with Fuchs and Peres when they claim that quantum theory fails to describe 
physical reality, and that the density operator captures everything that can meaningfully be said 
about a physical system. I sympathize with their revulsion against realistic construals of quantum 
states. The transmogrification of a probability algorithm into an evolving ontological state cannot fail 
to generate pseudoproblems and invite gratuitous solutions. But the claim that from the quantal 
probability assignments we cannot ``distill a model of a free-standing `reality' independent of our 
interventions'' is a {\it non sequitur\/} and a cop-out. I therefore also sympathize with those who 
carry on the search for such a model.

Physical science owes its immense success in large measure to its powerful ``sustaining 
myth''~\cite{MerminSM}---the belief that we can find out how things {\it really\/} are. Neither the 
ultraviolet catastrophe nor the spectacular failure of Rutherford's atomic model made physicists 
question their faith in what they can achieve. Instead, Planck and Bohr went on to discover the 
quantization of energy and angular momentum. If today we seem to have reason to question our 
sustaining myth, it ought to be taken as a sign that we are once again making the wrong assumptions, 
and it ought to spur us on to ferret them out. The Pondicherry interpretation does just that.

\section{\large Quantum mechanics, probabilities, and measurements}
The fundamental theoretical framework of contemporary physics, quantum mechanics, is a probability 
algorithm. This serves to assign, on the basis of outcomes of measurements that have been made, 
probabilities to the possible outcomes of measurements that may be made. The inevitable reference 
to ``measurement'' in all standard axiomatizations of unadulterated quantum mechanics was 
famously criticized by John Bell~\cite{Bell}: ``To restrict quantum mechanics to be exclusively about 
piddling laboratory operations is to betray the great enterprise.'' The search for more respectable 
ways of thinking about measurements began early. Since the discovery of special relativity in 1905, 
physicists had become used to calling them ``observations,'' and in 1939 London and Bauer 
\cite{LonBau} were the first to speak of ``the essential role played by the consciousness of the 
observer.''

Over the years, this red herring has taken many forms. To a few (e.g. \cite{Wigner}), it meant that 
the mind of the observer actively intervenes in the goings-on in the physical world, to some (e.g. 
\cite{dEspag,UlfBohr}), it meant that science concerns our perceptions rather than the goings-on 
``out there,'' while to most (e.g. \cite{FuPer,Petersen,Peierls}), it meant that quantum mechanics 
concerns knowledge or information about the physical world, rather than the physical world itself.

It is not hard to account for the relative popularity of the slogan ``quantum states are states of 
knowledge''~\cite{Fuchs}. The fundamental theory of the physical world is a probability algorithm, 
and there is a notion that probabilities are {\em inherently\/} subjective. Subjective probabilities are 
ignorance probabilities: they enter the picture when relevant facts are ignored. Because we lack the 
information needed to predict on which side a coin will fall, we assign a probability to each possibility. 
Subjective probabilities disappear when all relevant facts are taken into account (which in many cases 
is {\em practically\/} impossible).

The notion that probabilities are inherently subjective is a wholly classical idea. The very fact that our 
fundamental physical theory is a probability algorithm tells us that the probabilities it serves to assign 
are {\em objective\/}. The existence of objective probabilities (not to be confused with relative 
frequencies) is due to the fact that even the totality of previous measurement outcomes is insufficient 
for predicting subsequent measurement outcomes with certainty. The ``uncertainty principle''---the 
literal meaning of Heisenberg's original term, {\em fuzziness principle\/}, is more to the 
point---guarantees that, unlike subjective probabilities, quantum-mechanical probabilities cannot be 
made to disappear. If the relevant facts are sufficient to predict a position with certainty, there aren't 
any facts that would allow us to predict the corresponding momentum. What matters are facts, not 
what we can know about them.

Again, the very stability of matter hinges on the fuzziness principle. Ordinary objects have spatial 
extent, are composed of a (large but) finite number of objects without spatial extent, and neither 
collapse nor explode the moment they are created. The existence of such objects is made possible by 
the {\em objective fuzziness\/} of their internal relative positions and momenta \cite{Lieb}, not by 
our subjective uncertainty about the values of these ``observables.''

\begin{figure}
\begin{center}
\epsfig{file=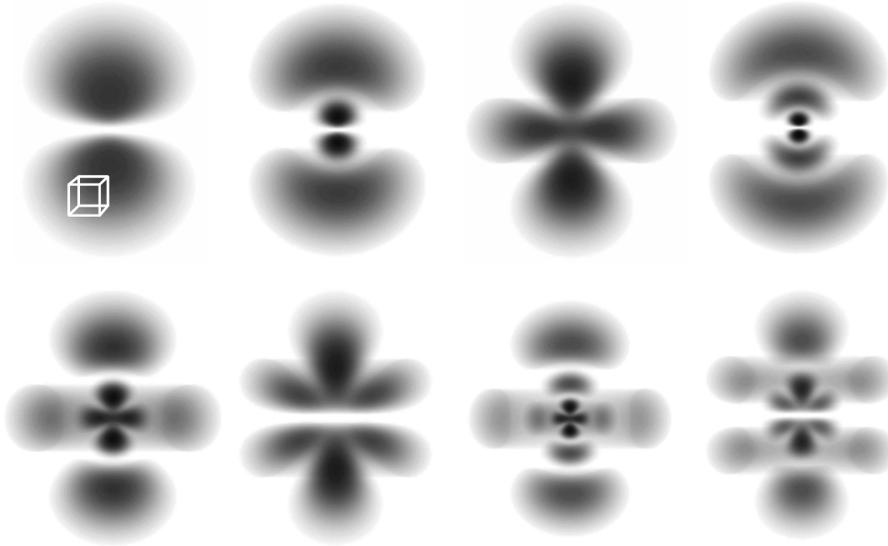,width=12cm}
\end{center}
\caption{The fuzzy position of the electron relative to the proton in various stationary states of 
atomic hydrogen.}
\end{figure}
What is the proper (i.e., mathematically rigorous and philosophical sound) way of describing the 
objective fuzziness of the quantum world? It is to assign {\em objective probabilities\/} to the 
possible outcomes of {\em measurements\/}. Take the familiar cloudlike images in Fig.~1. Each 
image represents the fuzzy position of the electron relative to the proton in a particular state of 
atomic hydrogen. Neither the electron nor the proton is shown. All we see is a fuzzy position. Or 
rather, all we see is a cloud of varying density, representing a continuous probability distribution: 
integrate the density of the cloud over a region like the box inserted in the first image, and get the 
probability of finding the electron inside if the appropriate measurement is made.

Now imagine that this measurement is actually made. It is an elementary measurement, in the sense 
that it answers a single yes/no question: is the electron inside that region? Before the measurement, 
the electron is neither inside nor outside, for if it were inside, the probability of finding it outside 
would be zero, as would the density of the cloud outside, and if it were outside, the probability of 
finding it inside would be zero, as would the density of the cloud inside. After the measurement, the 
electron is either inside or outside. In other words, the measurement creates a new state of affairs. If 
we want to describe a fuzzy state of affairs {\em as it is\/}, without messing with it, we must 
describe it counterfactually, by assigning probabilities to the possible outcomes of {\em 
unperformed\/} measurements. Clearly, quantum theory's inevitable reference to ``measurement'' 
has little to do with ``piddling laboratory operations.''

While the notion that probabilities are inherently subjective has induced many physicists to believe 
that quantum mechanics is an epistemic theory, the falsity of the notion that quantum mechanics is 
an epistemic theory has induced many other physicists to deny that the mathematical formalism of 
quantum mechanics is fundamentally and irreducibly a probability algorithm. This, too, is a {\em non 
sequitur\/} arising from the classical notion that probabilities are necessarily subjective. At present 
the physics community can be divided into three factions:

The first---the majority---doesn't care what (if anything) quantum mechanics is trying to tell us.

The second embraces agnosticism. It asserts that we cannot describe the quantum world as it is by 
itself; its features are forever beyond our ken. All we can usefully talk about is the statistical 
correlations between measurement outcomes (synchronic correlations between outcomes of 
measurements performed on different systems in spacelike relation as well as diachronic correlations 
between outcomes of measurements performed on the same system at different times).

The third insists that there must be a way of talking about the quantum world as it is by itself, 
independent of measurements. This faction is split into numerous sects, each declaring to see the 
light, the ultimate light. Go to any conference on quantum foundations, and you will find their priests 
pitted in holy war. (My thanks to C.~A. Fuchs~\cite{Fuchs1} for this observation.)

The agnostics and the priests both have a point and both are wrong. The agnostics have a point in 
that nothing of relevance can be said without reference to measurements. They are wrong in their 
belief that the features of the quantum world are beyond our ken. The priests have a point in that it is 
indeed possible to describe the features of the quantum world. They are wrong in their belief that 
these features can be described without reference to measurements. The objective fuzziness of the 
quantum world requires for its description assignments of probabilities to the possible outcomes of 
unperformed measurements.

\section{\large Ontological implications}
Consider again the unperformed elementary measurement designed to ascertain the electron's 
presence in, or absence from, the small boxlike region inside the first ``cloud.'' Before the 
measurement, the electron is neither inside nor outside this region. Yet being inside and being outside 
are the only relations that can hold between an electron and a given region. If neither relation holds, 
this region simply does not exist as far as the electron is concerned. But conceiving of a region~$R$ is 
tantamount to making the distinction between ``inside~$R$'' and ``outside~$R$.'' Hence we may 
say that the distinction we make between ``inside~$R$'' and ``outside~$R$'' is a distinction that 
the electron does not make. Or we may say that the distinction we make between ``the electron is 
inside~$R$'' and ``the electron is outside~$R$'' is a distinction that Nature does not make. It 
corresponds to nothing in the physical world. It exists solely in our heads.

This illustrates a general feature of the quantum world. It imposes limits on the distinctions that are 
physically warranted, not only spatial distinctions (between inside and outside) but also substantial 
distinctions (between this object and that object). Probabilities are calculated by summing over 
alternatives. Each alternative contributes an amplitude. In some cases we first square the amplitudes 
and then add the results. In all other cases we first add the amplitudes and then square the result. 
Whenever we add amplitudes first, the distinctions we make between the corresponding alternatives 
have no counterparts in the physical world.

Because of the limits that quantum mechanics imposes on the objectification of spatial distinctions, 
spatial distinctions are relative and contingent: {\em relative\/} because the difference between 
``inside~$R$'' and ``outside~$R$'' can be real for a given object at a given time and not have any 
reality for a different object at the same time or for the same object at a different time; and {\em 
contingent\/} because the reality of that difference (for a given object at a given time) depends on 
whether a relation (inside or outside) exists between this object and that region. Since no material 
object ever has an exact position, this in turn implies that the spatial differentiation of the physical 
world is incomplete. It doesn't go all the way down. If we mentally partition the world into smaller 
and smaller regions, there comes a point when there isn't any material object left for which these 
regions, or the corresponding distinctions, exist. And this again implies that a model of the quantum 
world cannot be built from the bottom up, on an intrinsically and completely differentiated spacetime 
manifold.

It is odd that for three quarters of a century, the ontological and/or epistemological status of the 
quantum-mechanical wave function~$\Psi$ has been the focus of a lively controversy, while the 
ontological status of the points and instants on which~$\Psi$ depends, has hardly ever been called 
into question. Virtually every published paper concerning the ontology of relativistic quantum 
mechanics (a.k.a. ``quantum field theory'') begins by postulating the existence of an intrinsically and 
completely differentiated spacetime manifold, despite the fact that quantum mechanics tells us that 
these points and instants exist solely in our heads. As long as the objective existence of such a 
manifold is postulated, it is safe to say that our attempts to beat sense into quantum mechanics are 
doomed.

The consequences of the limits that quantum mechanics imposes on our substantial distinctions, are 
not less drastic. We are allowed to distinguish between {\em this\/} particle and {\em that\/} 
particle only to the extent that particles have properties by which they can be distinguished, and they 
have such properties only to the extent that the possession of such properties can be inferred from 
the facts. As an illustration, consider an elastic scattering event involving two particles lacking 
properties that could serve as identity tags. Which outgoing particle is identical with which incoming 
particle? Quantum statistics implies that the question is meaningless. The challenge is to learn to 
think about the quantum world in ways that do not lead to meaningless questions. If we say that 
initially there are two incoming particles, one moving northward and one moving southward (say), 
and in the end there are two outgoing particles, one moving eastward and one moving westward 
(say), we cannot help asking that meaningless question. If, on the other hand, we say that initially 
there is one thing moving both northward and southward, and in the end there is one thing moving 
both eastward and westward, the meaningless question---which is which?---cannot be asked.

In a more philosophical language, what I am saying is that, in the quantum world, the concept of 
``substance'' betokens existence but it never betokens individuality. Individuality is strictly a matter 
of properties. If here you have a particle with these properties and there you have a particle with 
those properties, what you have is not two substances each with a set of properties but one 
substance with two sets of properties. This leads to the following conclusions:

(i)~Considered by themselves (out of relation to each other, out of relation to measurements, and out 
of relation to the laws that correlate measurement outcomes), the structureless constituents of 
matter are identical not merely in the weak sense of exact similarity but in the strong sense of {\em 
numerical\/} identity. (The evening star and the morning star are identical in this way.)

(ii)~Quantum mechanics is inconsistent with any attempt to construct a model of reality by 
assembling a multitude of distinct substances.

The quantum world thus can be built neither on the foundation of an intrinsically differentiated 
spacetime manifold nor out of a pre-existent multitude of individual substances. Any adequate model 
of the quantum world has to be constructed {\em from the top down\/}. What ultimately exists is 
{\em one\/}. Call it whatever you like. Matter and space both come into being when this enters into 
(more or less fuzzy) spatial relations with itself, for physical space is the totality of existing spatial 
relations (relative positions and relative orientations), while matter is the corresponding apparent 
multitude of relata---{\em apparent\/} because the relations are {\em self\/}-relations. This is 
about the simplest creation story that can be told, and it is a straightforward consequence of our 
fundamental theory of matter.

\section{\large The importance of measurements}
\label{SecImp}The existence of fuzzy variables calls for a criterion for the possession of a property 
(by a system), of a value (by a variable), or of a truth value (``true'' or ``false,'' by a proposition 
such as ``the electron is in~$R$ at the time~$t$''). There is a notion that probability~1 is 
sufficient for ``is'' or ``has.'' To see that this notion is wrong, let us 
ask why the probability of finding a given particle in the union~$C$ of two disjoint regions $A$ 
and~$B$, calculated according to the standard Born rule, is equal to the probability of finding the 
particle in~$A$ plus the probability of finding the particle in~$B$: $p(C)=p(A)+p(B)$. The answer 
would be self-evident if the particle's position were not fuzzy. In that case the particle would be 
either in~$A$ or in~$B$ whenever it is in~$C$, and the probabilities we calculate with the help of the 
quantum formalism would be subjective. To see that the answer is {\em not\/} self-evident, imagine 
two perfect (one hundred percent efficient) detectors $D(A)$ and $D(B)$ each monitoring one of 
those regions. If both $p(A)$ and $p(B)$ are greater than~0 (and therefore less than~1), it isn't 
certain that $D(A)$ will click and it isn't certain that $D(B)$ will click. Yet if $p(C)=1$ then it {\em 
is\/} certain that either $D(A)$ or $D(B)$ will click. How come?

The answer lies in the fact that quantum-mechanical probability assignments are invariably made on 
the (tacit) {\em assumption\/} that a measurement is successfully made: there is an outcome. The 
existence of an outcome means that either detector clicks. So there is no mystery here, but the upshot 
is that quantum mechanics gives us probabilities with which this or that outcome is obtained in a 
successful measurement, {\em not\/} probabilities with which this or that property, value, or truth 
value is possessed. Probability~1 is not sufficient for ``is'' or ``has.''

If probability~1 is not sufficient for ``is'' or ``has,'' then what is? As far as unadulterated, standard 
quantum mechanics is concerned---no surreal particle trajectories {\em\`a la\/} 
Bohm~\cite{Bohm}, no nonlinear modifications of the Schr\"odinger equation {\em\`a la\/} 
Ghirardi, Rimini and Weber~\cite{GRW} or Pearle~\cite{Pearle}, no extraneous axioms like the 
traditional eigenstate-eigenvalue link~\cite{vF} or the 
modal semantical rule~\cite{Dieks}---the only condition available is to be measured. To paraphrase a 
well-known dictum due to Wheeler~\cite{Wheeler}, no property (or value) is a possessed property 
unless it is an indicated property---unless, that is, its possession can be inferred from 
property-indicating events or states of affairs (a.k.a. ``measurements''). In other words, in the 
quantum world, properties and values are {\em extrinsic\/} rather than {\em intrinsic\/} (the latter 
word signifying ``possessed regardless of whether their possession is indicated'').

The fact that measurements {\em create\/} their outcomes was most forcefully brought home to us 
by Greenberger, Horne, and Zeilinger (GHZ)~\cite{GHZ}, who discovered a state of three entangled 
spin-$1/2$ particles that is an eigenstate of the following variables: the product 
$\sigma_x^1\sigma_x^2\sigma_x^3$ of the $x$~components of the three spins, and the products 
$\sigma_x^1\sigma_y^2\sigma_y^3$, $\sigma_y^1\sigma_x^2\sigma_y^3$, and 
$\sigma_y^1\sigma_y^2\sigma_x^3$ of the $x$~component of one spin and the $y$~components 
of the two other spins. The corresponding possessed eigenvalues are $-1$ for 
$\sigma_x^1\sigma_x^2\sigma_x^3$ and $+1$ for the remaining three products. If the individual 
spin components were in possession of either of their possible values ($+1$ or $-1$) regardless of 
measurements, the following equations would hold for their possessed values
\be
s_x^1s_y^2s_y^3=1,\quad s_y^1s_x^2s_y^3=1,\quad s_y^1s_y^2s_x^3=1,\quad 
s_x^1s_x^2s_x^3=-1.
\ee
Multiply the left-hand sides of the first three equations to find that it equals $s_x^1s_x^2s_x^3$. 
The product of the right-hand sides equals~1, implying that $s_x^1s_x^2s_x^3=1$, in direct 
contradiction to the fourth equation. So unless a (compatible) set of spin components are actually 
measured, no values can be attributed to them.

Why is it that physical variables have values only if, and only to the extent that, values are indicated 
(``measured'')? Take positions. Since spatial distinctions are relative and contingent, physical space 
is anything but an intrinsically differentiated expanse. Neither regions of space nor positions exist 
``by themselves.'' Philosophically speaking, positions are properties, not substances; they exist only 
if and when they are possessed. But properties are possessed only if their possession is indicated, and 
the possession of the property of being in a region~$R$ can only be indicated if~$R$, or the 
distinction between inside~$R$ and outside~$R$, is realized (made real) by a detector (in the 
broadest sense of the word: anything capable of indicating the presence of something somewhere). A 
detector, therefore, performs {\em two\/} necessary functions: to indicate the presence of an object 
in its sensitive region~$R$, and to make the predicates ``inside~$R$'' and ``outside~$R$'' available 
for attribution. Much the same goes for any other measurement apparatus, not least because every 
measurement outcome is ultimately indicated by the position of something like a pointer: it serves to 
realize indicatable properties as well as to indicate them.

And why are certain variables incompatible? Because the corresponding properties cannot be 
simultaneously {\it realized\/}. This is particularly clear in the case of spin components. In the 
absence of an axis that is physically realized by a magnetic gradient, the values ``up'' and ``down'' 
are undefined. Since the superposition of two magnetic fields with different gradients yields a single 
magnetic field with a single gradient, only one axis can be realized at a time.

If the points and instants on which~$\Psi$ functionally depends have no counterparts in the physical 
world, what is the meaning of the spatial and temporal arguments of~$\Psi$? The wave function is 
one of the quantum-mechanical tools for calculating probabilities. It is the appropriate tool when we 
assign probabilities on the basis of an earlier outcome of a complete measurement. We provide two 
kinds of information: (i)~the difference $t-t'$ between the {\em physically realized\/} time~$t$ of 
the measurement to the possible outcomes of which probabilities are assigned, and the {\em 
physically realized\/} time~$t'$ of the measurement on the basis of whose outcome probabilities are 
assigned; (ii)~a {\em physically realized\/} region~$R$ (the small boxlike region above, or any 
region of the system's configuration space). Without this information, the wave function is an empty 
shell devoid of physical significance.

Having plugged in this information, $\Psi$~gives us the probability of finding the electron (or the 
system) in $R$ {\em given that\/} the corresponding elementary test is successfully made at the 
time~$t$. Once we have this probability for all conceivable regions, we can calculate the mean value, 
the variance, and the higher moments that characterize a fuzzy relative position. Using the Fourier 
transform of~$\Psi$, we can similarly compute the various moments of the corresponding fuzzy 
momentum. And under certain conditions this gives us a complete description of a fuzzy state of 
affairs. (The complete description of the fuzzy state of affairs that obtains {\em between\/} 
successive measurements is given by the time-symmetric probability algorithm due to Aharonov, 
Bergmann, and Lebowitz~\cite{ABL}, rather than by the Born rule, which bases probabilities on 
earlier {\em or\/} later outcomes~\cite{MohrhoffABL}.)

\section{\large Explanations or delusions?}
Consider once again three spin-$1/2$ particles in the GHZ state. By measuring the $x$~components 
or the $y$~components of two spins, we can predict with certainty the $x$~component of the third 
spin. By measuring the $x$~component of one spin and the $y$~component of another spin, we can 
predict with certainty the $y$~component of the third spin. And by measuring the $z$~component of 
one spin, we can predict with certainty the $z$~components of the two other spins. How is it possible 
to predict any spin component of any of the three particles after subjecting the other particles to the 
appropriate measurements, considering (i)~that the GHZ correlations are independent of the distance 
between the three particles (in principle they can be light years apart), and (ii)~that these 
measurements do not reveal pre-existent values but {\em create\/} their outcomes?

Counter question: Why do the GHZ correlations {\em seem\/} impossible? If we believe, as Einstein 
did, that ``things claim an existence independent of one another'' whenever they ``lie in different 
parts of space''~\cite{Einstein}, then such correlations are indeed impossible. Fact is that the three 
particles, irrespective of the distances between them, are {\em not\/} independent of one another. 
Fiction is that they lie in different parts of space. As we have seen, space has no parts. If we insist on 
thinking of space as a self-existent expanse, to which spatial relations owe their quality of spatial 
extension, quantum mechanics does not permit us to think of this expanse as being divided. The 
spatial multiplicity of the world rests on the existence of more or less fuzzy relations, rather than on 
the existence of spatial parts. One might say, paradoxically yet to the point, that, as far as ``space 
itself'' is concerned, there is only one place, and this is everywhere. Instead of separating things, 
space (qua expanse) unites them by being devoid of any kind of multiplicity. However, since being 
extended and being undifferentiated makes a rather paradoxical combination of properties, it is much 
better to look on space, not as a self-existent expanse, but as the totality of spatial relations that hold 
between material objects. Then one cannot even conceive of ``parts of space.''

We saw, moreover, that the quantum world has room for only one substance. Considered by 
themselves, the structureless constituents of matter are identical in the strong sense of numerical 
identity. All existing relations are self-relations. How then could things possibly ``claim an existence 
independent of one another''?

It is one thing to dispose of misconceptions that make the peculiar nonlocal behavior of quantum 
systems seem impossible. It is quite another to {\em explain\/} the quantum-mechanical correlation 
laws. If these laws are indeed the fundamental laws of physics (and apart from our dogged insistence 
on explaining from the bottom up, we have no reason to believe that they are not), then they cannot 
be explained the way Kepler's laws of planetary motion can be explained by Newton's law of gravity. 
Only a law that is not fundamental can be so explained.

Can we interpret the quantum-mechanical correlation laws as {\em descriptive\/} of a physical 
mechanism or process? Where the synchronic correlations are concerned, this notion is patently 
absurd. Alas, to many the absurdity of interpreting the diachronic correlation laws as descriptive of 
some physical process does not seem to be obvious, despite the preposterous consequences of doing 
so, such as wave function collapse or lack of relativistic covariance and of gauge invariance.

As an algorithm for assigning probabilities to possible measurement outcomes on the basis of actual 
measurement outcomes, $\Psi$~has two obvious dependences. It depends continuously on the time 
of a measurement: if you change this time by a small amount, the probabilities assigned to the 
possible outcomes change by small amounts. And it depends discontinuously on the outcomes that 
constitute the assignment basis: if you take into account an outcome that was not previously taken 
into account, the assignment basis changes unpredictably as a matter of course. Transmogrify~$\Psi$ 
into the description of an evolving, instantaneous state of affairs, and you are faced with the mother 
of all quantum-mechanical pseudoproblems: why are there two modes of evolution rather than one?

While the real trouble with von Neumann's formulation of quantum mechanics~\cite{vN} is that it 
postulates two modes of evolution rather than {\em none\/}, many believe that the trouble with it is 
that it postulates two modes of evolution rather than {\em one\/}. Instead of addressing the root of 
the disease---the transmogrification of a probability algorithm into an evolving ontological 
state---they make matters worse by postulating a universal quantum state that evolves 
deterministically at all times.

This postulate is inconsistent with the ontological implications of the quantal probability algorithm. If 
the wave function represented something that evolves deterministically, it would evolve in a 
completely differentiated time. Unitary evolution implies that the world is infinitely differentiated 
timewise, whereas the quantum-mechanical correlation laws imply that the world is infinitely 
differentiated neither spacewise nor timewise, as the following will show.

Like the properties or values themselves, the times at which properties or values are possessed must 
be indicated (measured) in order to exist. As detectors are needed not only to indicate but also to 
realize positions, so clocks are needed not only to indicate but also to make times available for 
attribution. Since clocks realize times by the positions of their hands, and since exact positions do not 
exist, neither do exact times. (Digital clocks indicate times by transitions from one reading to another, 
without hands, but the ``uncertainty'' principle for energy and time implies that these transitions 
cannot occur at exact times.~\cite{Hilge}) Exact times therefore are not available for attribution. 
Like the existing spatial relations, the existing temporal relations are fuzzy. From this the incomplete 
temporal differentiation of the physical world follows in exactly the same way as its incomplete 
spatial differentiation follows from the fuzziness of positions.

There are several reasons, most of them psychological and none of them physically warranted, why 
we tend to believe in the existence of an evolving instantaneous physical state. And once we believe 
that there is such a state, it is obviously hard to avoid thinking of the wave function as representing 
such a state, and of the evolution of the wave function as a physical process.

For one, classical physics appears to be consistent with these notions. It therefore deserves to be 
pointed out that even the idea that a {\em classical\/} dynamical law describes (i)~the evolution of 
a physical state and (ii)~a physical {\em process\/}, rests on nothing but the physically 
unwarranted transmogrification of an algorithm for calculating the {\em effects\/} of interactions 
into a physical process. Take classical electrodynamics. It allows us to calculate the effects that 
charges have on charges. The calculation involves two steps. (i)~Given the distribution and motion of 
charges, we calculate six functions, the components of the electromagnetic field. (ii)~Given these six 
functions, we calculate the effect that those charges have on any other charge. And that's that. The 
rest is embroidery, including the belief that the unobservable electromagnetic field is a physical entity 
in its own right, that it is locally generated by charges, that it locally acts on charges, that it physically 
mediates interactions and propagates energy and momentum (which are construed as some 
localizable kind of stuff), and that this {\em explains\/} how charges act on charges. If we were 
honest, we would admit that all we have in classical physics is correlations that, being deterministic, 
can be thought of as correlations between causes and effects. We do not have the merest notion of 
how (through what mechanism or process) causes produce effects.

Among the psychological reasons for our belief in an evolving instantaneous physical state and the 
ensuing misconceptions, the following deserves to be mentioned: while our successive experience of 
reality makes it natural for us to hold that only the present is real, or that it is somehow ``more real'' 
than the future or the past, our self-experience as agents makes it natural for us to hold that the 
known or in principle knowable past is ``fixed and settled,'' and that only the unknown and 
apparently unknowable future is ``open.'' None of this has anything to do with physics.

For one thing, it is impossible to consistently project the experiential Now into the physical world. To 
philosophers, the perplexities and absurdities entailed by the notion of a changing objective present 
or a flowing time are well known~\cite{Audi}. To physicists, the subjectivity of a temporally 
unextended yet persistent and persistently changing present was brought home by the relativity of 
simultaneity. The same is implied by quantum mechanics, inasmuch as the incomplete differentiation 
of the quantum world rules out the existence of an (evolving) instantaneous state.

For another thing, the physical correlation laws (whether classical or quantum) know nothing of a 
preferred direction of causality. They are time-symmetric. They let us retrodict as well as predict. The 
figment of a causal arrow is a projection, into the physical world, of our sense of agency, our ability to 
know the past, and our inability to know the future. It leads to the well-known folk tale according to 
which causal influences reach from the nonexistent past to the nonexistent future through persisting 
``imprints'' on the present: If the past and the future are unreal, the past can influence the future 
only through the mediation of something that persists. Causal influences reach from the past into the 
future by being ``carried through time'' by something that ``stays in the present.'' This evolving 
instantaneous state includes not only all presently possessed properties but also traces of everything 
in the past that is causally relevant to the future.

In classical physics, this is how we come to conceive of fields of force that evolve in time (and 
therefore, in a relativistic world, according to the principle of local action), and that mediate between 
the past and the future (and therefore, in a relativistic world, between local causes and their distant 
effects). In quantum physics, this is how we come to seize on a probability algorithm that depends on 
the relative time between measurements and on the outcomes of earlier measurements, to 
transmogrify the same into an evolving instantaneous state, and to think of its evolution as a physical 
process.

The bottom line: we have no reason to mourn the loss of our ability to interpret the physical 
correlation laws as descriptive of a physical process (the evolution of a physical state), inasmuch as 
the idea that we once had this ability is a delusion. We have lost nothing. Instead, we have gained 
significant insights into the spatiotemporal and substantial aspects of our world.

\section{\large Quantum mechanics and reality}
The extrinsic nature of physical properties and values appears to entail a vicious regress. No value is a 
possessed value unless it is indicated, and a pointer position is no exception; it has a value only 
because, and only to the extent that, its value is indicated by other ``pointer positions.'' How, then, 
can a detector (in the broadest sense) realize a region~$R$ (or the distinction between 
``inside~$R$'' and ``outside~$R$'')? Is it possible to terminate this regress without invoking an 
extra-physical principle like consciousness? Indeed it is. As we shall see, this possibility crucially 
depends on the incomplete spatial differentiation of the physical world.

The possibility of obtaining evidence of the departure of an object~$O$ from its classically predictable 
position (given all relevant earlier position-indicating events) calls for detectors whose position 
probability distributions are narrower than~$O$'s. Such detectors do not exist for all objects. Some 
objects have the sharpest positions in existence. For these objects, the probability of obtaining such 
evidence is extremely low. Hence {\em among\/} these objects there are many of which the 
following is true: every one of their indicated positions is consistent with (i)~every prediction that is 
based on their previous indicated positions and (ii)~a classical law of motion. Such objects deserve to 
be called ``macroscopic.'' To enable a macroscopic object to indicate an unpredictable value, one 
exception has to be made: its position may change unpredictably if and when it serves to indicate 
such a value.

Decoherence investigations~\cite{Zurek2003, Blanchard2000, Joos2003} have shown for various 
reasonable definitions of ``macroscopic'' that the probability of finding a macroscopic object where 
classically it could not be, is extremely low. This guarantees the existence of macroscopic objects 
according to our stricter definition, which are never actually found where classically they could not be. 
The correlations between the indicated positions of these objects are deterministic in the following 
sense: their fuzziness never evinces itself through outcomes that are inconsistent with predictions 
that are based on earlier outcomes and a classical law of motion. Macroscopic objects (including 
pointers) follow trajectories that are only counterfactually fuzzy. That is, they are fuzzy only in 
relation to an imaginary background that is more differentiated spacewise than is the actual world. In 
the physical world, there is nothing over which they are ``smeared out.'' So we cannot say that they 
are fuzzy---nor can we say that they are sharp: ``not fuzzy'' implies ``sharp'' only if we postulate the 
intrinsically and completely differentiated background space that does not exist in the quantum world.

To terminate that seemingly vicious regress, we must be allowed to look upon the positions of 
macroscopic objects---macroscopic positions, for short---as intrinsic, as self-indicating, or as real {\it 
per se\/}. The reason why this is indeed legitimate, is that the extrinsic nature of the values of 
physical variables is a consequence of their fuzziness. If macroscopic positions are not manifestly 
fuzzy, we have every right to consider them intrinsic---notwithstanding that they are at the same time 
extrinsic, for even the Moon has a position only because of the myriad of ``pointer positions'' that 
betoken its whereabouts. The reason why macroscopic positions can be both extrinsic and intrinsic is 
that they indicate each other so abundantly, so persistently, and so sharply that they are only 
counterfactually fuzzy.

Nothing therefore stands in the way of attributing to the entire system of macroscopic positions an 
independent reality, and nothing prevents us from considering the entire system of possessed relative 
positions or spatial relations (including the corresponding multitude of material relata) as 
self-contained. What the extrinsic nature of physical properties forbids us, is to attribute independent 
reality to an individual macroscopic position, and to model the physical world from the bottom up. We 
are not entitled to the belief that the macroworld is what it is because its constituents are what they 
are. The ``foundation'' is the macroworld---the system of macroscopic positions, in which 
value-indicating events occur as unpredictable transitions in compliance with the 
quantum-mechanical correlation laws. All other properties of the physical world (in particular the 
properties of the so-called ``quantum domain'') exist solely because they are indicated by the 
goings-on in the macroworld (the so-called ``classical domain''). As philosophers would say, the 
properties of the quantum domain {\it supervene\/} on the goings-on in the classical domain.

A fundamental physical theory concerned with nothing but statistical correlations between 
value-indicating events, presupposes the occurrence of such events. The philosphically most 
challenging problem is to understand how such a theory can be complete---how it can at the same 
time encompass the value-indicating events. The solution of this problem calls for a judicious choice 
on our part: which substructure of the entire theoretical structure of quantum mechanics corresponds 
to what exists? Independent reality can be attributed consistently neither to an intrinsically divided 
``manifold'' nor to a multitude of microscopic constituents nor to an evolving instantaneous 
``quantum state'' but only to the macroworld.

The problem of making sense of quantum mechanics is often misconceived as the problem of how a 
classical domain emerges in a quantum world. Some feel called upon to explain how possibilities---or 
worse, probabilities~\cite{Treiman}---become facts, some try to show how properties 
emerge~\cite{JoosZeh}, and some wish to tell us why events occur~\cite{Pearle2}. Since the 
quantum domain supervenes on the goings-on in the macroworld, and since independent reality can 
be consistently attributed only to the macroworld, there is no ``underlying reality'' from which the 
macroworld could emerge. Saying in common language that a possibility becomes a fact is the same 
as saying that something that is possible---something that {\it can\/} be a fact---actually {\it is\/} 
a fact. This non-problem becomes a pseudoproblem if the common-language ``existence'' of a 
possibility is construed as another kind of existence---a matrix of ``propensities''~\cite{Popper} or 
``potentialities''~\cite{Heisenberg, Shimony}---that transforms into the genuine article 
(nonexistence proper or existence proper) by way of measurement. (Pseudoproblems of this kind are 
bound to arise if one misconstrues a probability algorithm as an evolving, instantaneous physical 
state.) There is no such matrix.

Nor is it possible to explain why events occur. As we observed in Sec.~\ref{SecImp}, 
quantum-mechanical probability assignments are made with the tacit assumption that a measurement 
is successfully made: there is an outcome. If quantum mechanics is our fundamental theoretical 
framework, and if every quantum-mechanical probability assignment {\it presupposes\/} the (actual 
or counterfactual) success of a measurement (in the form of a value-indicating event), then quantum 
mechanics cannot possibly supply sufficient conditions for the occurrence of such an event. The 
events by which properties or values are indicated, are {\it uncaused\/}. (While this implies that 
perfect detectors are a fiction, it does not prevent us from invoking perfect detectors to assign 
probabilities to the possible outcomes of {\it unperformed\/} measurements.)

So does quantum mechanics encompass the system of macroscopic positions, in which 
value-indicating events occur as unpredictable transitions? It does. The fuzziness of all existing 
relative positions implies the incomplete spatial differentiation of the physical world, on account of 
which the least fuzzy positions in existence are at the same time sharp and the least fuzzy extrinsic 
properties are at the same time intrinsic. This makes the attribution of independent reality to the 
system of macroscopic positions both possible and inescapable.

Do the quantum-mechanical correlation laws {\it account\/} for the correlata which they 
presuppose? Since value-indicating events are uncaused, they don't. This is no more a shortcoming of 
quantum physics than it is a shortcoming of classical physics that it cannot explain why there is 
anything, rather than nothing at all. Physics is concerned with laws, and hence with nomologically 
possible worlds---worlds consistent with the laws. It would be preposterous to expect from it an 
explanation of why the actual world exists. Quantum physics is concerned with correlation laws, and 
it would be preposterous to expect it to account for the existence of the correlata. We have done our 
utmost if we have demonstrated not only the consistency of the correlation laws with the existence of 
correlata (which is trivial) but also the {\it completeness\/} of the correlation laws: they can account 
for anything but the factuality of (value-indicating) facts. Many aspects of a value-indicating event 
can be understood in terms of the deterministic correlations that structure the macroworld, and 
everything else {\it but\/} the actual occurrence of such an event can be understood in terms of the 
statistical correlations that structure the (supervenient) microworld. The absence of causally 
sufficient conditions for the occurrence of an event of type~X does not preclude a complete 
theoretical analysis of events of type~X. As we have seen, our inability to formulate sufficient 
conditions for the occurrence of a value-indicating event does not imply the incompleteness of 
quantum mechanics but is instead a consequence of the incompleteness of the quantum world, in 
particular its incomplete spatiotemporal differentiation.

\end{document}